# The ethics of artificial intelligence in the life sciences: Universality, cultural diversity and an architecture of care

Jean-Pierre Changeux[1], Gustavo Deco[2-4] and Morten L. Kringelbach[4-7]

1. Neuroscience Department, Institut Pasteur, Collège de France, Paris F-75005, France.
2. Center for Brain and Cognition, Computational Neuroscience Group, Faculty of Medicine and Life Sciences, Universitat Pompeu Fabra, Barcelona, Spain
3. Institució Catalana de la Recerca i Estudis Avançats (ICREA), Barcelona, Spain
4. International Centre for Flourishing, Universities of Oxford (UK), Aarhus (Denmark) and Pompeu Fabra (Spain)
5. Centre for Eudaimonia and Human Flourishing, Linacre College, University of Oxford, Oxford, UK
6. Department of Psychiatry, University of Oxford, Oxford, UK
7. Center for Music in the Brain, Department of Clinical Medicine, Aarhus University, Aarhus, DK.

The life sciences and health research have started to benefit from artificial intelligence, which raises ethical concerns that are real but, we argue, not special. Any science should be governed by values that rest on how the human brain is built and socialised rather than anything distinct to artificial intelligence. Importantly, the human brain has a different, much less costly computational architecture than these machines. This is achieved through the orchestration of a global neuronal workspace, and through reward best described not as a quantity to be maximised but as a continuous cycle of wanting, liking and satiety. As such, this creates the deep tension running through the ethics of the human person, between the universality of ethical judgement and the diversity of morals. The brain networks of the global workspace and emotion are universally shared, but the diversity of content is shaped by epigenetic appropriation of the particulars of the physical, social and cultural world, which makes every person unique. Still, if we were to build machines on these principles rather than the present unaffordable reward maximisers, the question of their governance would change from restraint to upbringing. We set out the institutions such a future would require, together with the questions that remain open.

*"Thought and sensory perception are merely changes in matter, and therein lies an arbitrary limit to the truth… being a product of brain matter, the best mathematical model shall never give a complete and exhaustive description of reality" – inspired by fragments of Democritus (Voilquin, 1964)*

Over a very short period of time, the life sciences and health research have started to become transformed by artificial intelligence, from generation of texts to medical advice and folding of proteins. The rise of the machines has created deep ethical concerns, over a wide range of issues including the loss of employment, threats to human rights and degradation of the climate. Invariably these fall hardest on already marginalised groups leading to concentration and instability of our political systems. Some may therefore conclude that this requires the development of a new field of the ethics of artificial intelligence.

However, we take a contrasting view, arguing here that artificial intelligence is no different from any other product of science; it is not a new kind of moral agent since it does not possess free ethical judgment and does not define its own goals nor weigh the consequences of their use. Our arguments is aligned with Georges Canguilhem (1977) who insisted that to understand a piece of scientific knowledge, we must know its origin, the conditions of its production and its destination. Ultimately, these are therefore questions about human beings and their institutions rather than about the instruments and the ethics for artificial intelligence is the same ethics needed for all science. There is, however, an urgency to this question linked to the strain on the very human faculties that would govern it. We argue that these faculties depend on the building, growing and socialisation of the human brain and a better understanding of this changes what we must ask of the machines, both current and those that may be built in future.

Here, we first compare human and artificial intelligence in terms of computation and set out why the brain is a particular kind of computer with reward and emotion at its centre. Second, we follow the ethics of the human person as it has been drawn by leading thinkers such as Kropotkin, Ricoeur and Kant. We show that the architecture of the brain supplies mechanisms underlying the central tension between a universal capacity for ethical judgement and a deep diversity of morals across persons and cultures. In conclusion, while a new ethics is not needed for artificial intelligence, we point to necessary safeguards and institutions needed.

### *Comparing human and artificial intelligence*

Artificial intelligence has started to match human performance on various tasks while also being a powerful tool to use machine learning on large collections of data to pair problems with their solutions. As an example, AlphaFold can predict the three-dimensional structure of a protein from its amino-acid sequence across more than two hundred million entries (Jumper *et al.*, 2021; Varadi *et al.*, 2022). It does so from learning existing structures, but returns no new information about the kinetics of folding and, crucially, misses essential features of protein function such as allostery (Changeux, 2012). Equally, large language models have powerfully entered our everyday lives, answering almost any question immediately, yet this comes at the price of frequent errors and confabulations, sometimes leading to an illusion of understanding. In some ways they are comparable to what Kahneman and Tversky described as a first system of rapid judgement running but without the second, deliberate system (Kahneman, 2011; Tversky and Kahneman, 1974). Still, this computation requires an enormous energy budget where a single training run can draw on the order of a gigawatt-hour and the deployed model spends again at every query. This is unsustainable for a planet with limited energy.

The human brain functions very differently, running on about twenty watts, near the food energy of four bananas across a day. On this modest budget the human brain keeps a body alive among hunger, thirst, warmth and danger that rise and fall over time, often unpredictably. Still, the human brain holds a model of the world together and learns without pause. The human brain is a self-organising system with many nested levels, from atoms and molecules through neurons and local circuits to networks, to conscious processing, and outward to language, social life and culture (Changeux, 2017). This allows for an inferential system that anticipates and evaluates rather than just reacting. The regulation of this system is multi-level, where especially the molecular level matters because it is where pharmacological agents act and where mental illness may arise.

Artificial intelligence is organised differently, carrying out gigantic digital and parallel computations by formal algorithmic routes on electronic devices. Importantly, they share none of the biological molecular elements of the brain and therefore do not have similar pharmacology nor human-like mental health problems. They are asked to solve very specific questions using as much time and power as we can afford.

The brain is fundamentally different in that it must discover for themselves what matters, choose continually, quickly and on very little energy. This is achieved through the compression of the vast field of what might be done into a simple orientation, positive or negative, that disposes the organism to approach or to avoid. This fast and unconscious compression of the survival decision space is what we call reward or *emotion*. The brain uses this heuristic to arrive at particular answers through a process where its own physical activity settles into a pattern rather than by working through possibilities in sequence. This is a key mechanism that makes brains at once fast and cheap (Kringelbach *et al.*, 2026). The processing ignites the Global Neuronal Workspace (Baars, 1988; Dehaene *et al.*, 1998; Mashour *et al.*, 2020), where the information can be held, compared and reconsidered before anything is done, and then broadcast to the rest of the brain (Baars, 1988). The Global Neuronal Workspace is therefore a seat of deliberation and of ethical judgement.

Importantly, reward and emotion in the brain are not quantities to be maximised. This is different from the long history of reward as a single value that the brain estimates and pursues, which is how a great deal of artificial intelligence is built. Instead, the study of the brain has shown that reward consists of a continuous cycle of wanting, liking and satiety with distinct machinery (Berridge and Robinson, 1998; Berridge *et al.*, 2009; Kringelbach and Berridge, 2009). *Wanting* draws the animal towards a goal and acts as a forecast of what is to come. *Liking* is the read-out of the reward that actually arrives, and satiety is the forecast of repletion that closes the episode and settles when the goal becomes eligible again. Learning runs throughout, and the three phases turn as a cycle of emotion, renewing itself and holding the whole system far from equilibrium and near the point of criticality where the brain works.

The potential for survival arises when approach and avoidance are anchored in this way to the condition of a body, and when the individual can therefore be satisfied and deprived. This is not available to a merely maximising system. Emotion of this kind can also be read from outside, as the response of adults to the sight and sound of infants makes plain (Kringelbach *et al.*, 2008; Kringelbach *et al.*, 2016; Lorenz, 1943). This legibility is what allows one individual to be moved by the state of another, giving a physical form to what Kropotkin called mutual aid and to what Ricoeur called holding oneself as another. This evolutionary scheme shows that care is not a sentiment laid over a computation, nor benevolence added afterwards as a policy, but the coupling of two individuals through their mutual emotion.

Seen this way, the large language model and the world model are easy to place (Kringelbach *et al.*, 2026). They are trained to predict the output of multilevel human higher brain functions (Changeux, 2017), which organised around emotion as we just saw. Equally, the large language model has been shown to carry the trace of that organisation in everything it produces and can speak of care and of value fluently and at length (Sofroniew *et al.*, 2026). But they only carry the trace without the substance, being strictly functionalist models with no energy budget, no pressure to stay far from equilibrium and above having all no reward cycle of their own, no wanting that can be satiated and no true liking anchor. Such a system can be said to be a shadow of the human brain architecture that produced its training material, and crucially a shadow traces the shape of what casts it without possessing any of its weight.

### *Ethics as a social relationship and the underlying brain architecture*

At its centre, ethics concerns human social relationships and, at bottom, the survival of the species through mutual aid. Kropotkin argued that nature is the first teacher of ethics and that the moral sense, innate in humans as in social animals, climbs through successive steps, from sociability and mutual aid, through sympathy and good will with their rules of justice and equality, to generosity as the highest reach of moral evolution (Kropotkin, 1902). Ricoeur gave the ethical aim its most compact form as a "good life with and for others in just institutions", and named the capacity at its heart as the ability to hold oneself as another (Ricoeur, 1990). This is close to what Mauss had described as the principle of mutual benefit (Mauss, 1925). Kant placed the dignity and the autonomy of the person above all price (Kant, 1785), and with Rousseau (1762) and Thoreau (1854) we would add respect for the natural environment on which the person depends. Spinoza saw the root of the matter when he wrote that we should consider human actions and appetites as if it were a question of lines, planes and bodies (Spinoza, 1677), which in the language of neuroscience would mean that people judge things according to the organisation of their brain.

The brain architecture described above provides the tradition with a mechanism, since the mutual aid that Kropotkin places at the root is, in physical terms, the coupling of two systems through their legible emotion. The clearest evidence of that legibility is the *kindchenschema* or *infant schema*, the configuration of large eyes and rounded form that reliably moves an adult of almost any mammal to protect and to nurture (Kringelbach *et al.*, 2008; Kringelbach *et al.*, 2016; Lorenz, 1943). It works because one animal can read and be moved by the valence of another. Care, on this account, is not an emotion laid over a computation. It is the architecture of weighing acceptable against unacceptable in a way other people can read and answer, and it is precisely the architecture that a single maximised number cannot have. Sympathy and empathy are the names this coupling takes in the human case, and they rest on the same legible emotion.

Two intrinsic difficulties stand in the way, and both are recognisable in the same terms. Later in life, Lorenz described the human being as caught in an evolutionary disharmony, where we gave priority to the cognitive faculties rather than the social components of empathy and sympathy, and from this imbalance inter-individual violence emerges (Lorenz, 1963). Similarly, Günther Anders wrote in his book *The Obsolescence of Man* of a second and widening gap, between our capacity to create science and new technologies and our capacity to conceive the consequences of that creation, resulting in a "blindness to the apocalypse" (Anders, 1956). In the language of the brain architecture, the first difficulty is the lag of the slow social extension of emotion behind its fast core of reward, and the second is a failure of the very deliberation that the global workspace exists to perform. Ethics is required precisely because these gaps are real, and because the survival of the species depends on closing them.

### *Ethical universalism and the universal brain architecture*

Ethics is essentially perched between universalism and diversity. Is ethical judgement universal across humans or relative to moral and cultural setting. Take Ricoeur's capable person who is a rational and conscious, engaged in social relationships with a personal identity and a social status (Ricoeur, 1990). This person communicates intentionally, attends and feels, represents the world, recognises themselves, distinguishes themselves from others and searches for mutual recognition and shared reward.

The evidence from the brain points first to a universal floor. Watching another person in pain has been shown to provide activity in the observer in the affective regions engaged by pain itself, whether the pain is perceived or only imagined (Singer *et al.*, 2004). Similarly, damage to the orbitofrontal cortex can produce emotional disturbances such as aggressivity and a lack of remorse while the capacity to attribute states to others is preserved, so that the inhibition of violence is selectively lost (Raine *et al.*, 1997).

The universality reaches deeper than any single circuit, into the mode of conscious processing itself. The theory of the Global Neuronal Workspace proposes a neuronal architecture with long-range axons from cortical pyramidal cells interconnecting and reciprocally broadcasting signals across a fronto-parieto-temporal-cingulate network. When a content is amplified into this network it is ignited into the global neuronal workspace and broadcast at once to memory, attention, evaluation and action (Dehaene and Changeux, 2011; Dehaene *et al.*, 1998; Mashour *et al.*, 2020). In its original conception (1998), the Global Neuronal Workspace theory spans several levels of organisation and includes explicit implementation from molecules, neural circuit, and global modulation up to cognition. The workspace acts as a temporal buffer between past events and future actions, and in doing so it contributes to decision-making, to self-evaluation and to autonomy as well as to their pathology and pharmacology. This corresponds to how Ricoeur described “consciousness”, as a space for deliberation on thought experiments in which moral judgement is exercised in a hypothetical manner.

The Global Neuronal Workspace is nevertheless where the brain architecture and the ethics become one. The broadcast does two things at once, being the step at which the outputs of emotion become available to the system as a whole, which is where feeling may arise, so that the theory of conscious access and the line we draw between emotion and feeling turn out to describe a single mechanism seen from two sides. This has been described by the *entangled loop*, a related theory of consciousness based on a trifold of emotion, interference and hierarchy (Kringelbach *et al.*, 2026). It is also the space in which a content can be held, set beside others and turned over before anything is done about it, which is the self-evaluation that Ricoeur called deliberation. The seat of conscious processing is therefore also the seat of moral deliberation. This universal form is the common *Humanitas* that Kant proposed, and it is the floor on which any cooperation across difference must stand (Changeux, 1997).

Artificial intelligence is generally assumed not to have a Global Neuronal Workspace, but there have been recent claims of this in Claude, the language model (Gurnee *et al.*, 2026). We are preparing a separate note on this important issue, where the main conclusion is that actual multilevel conscious processing proposed by the Global Neuronal Workspace theory should not, in any instance, be confused with any strictly functionalist description by artificial intelligence.

***Diversity of morals and epigenetic learning***

Importantly, moral diversity arises from how the universal brain architecture is not closed at birth but continuously engaged with the world. The brain multiplies its weight fivefold from newborn to adult, and its circuits are shaped by the selective stabilisation of synapses under the pressure of activity and experience (Changeux and Danchin, 1976; Edelman, 1978). The variability theorem shows that different connective organisations can support the same behavioural abilities, so that the same function is reachable by many wirings (Levinthal *et al.*, 1976). Spontaneous activity organises circuits before experience begins (Sretavan *et al.*, 1988), and the acquisition of a cultural skill leaves its mark in the connectivity of the brain, as the comparison of literate and illiterate brains during reading makes plain (Dehaene *et al.*, 2010). The physical, social and cultural environment is imprinted in the brain through synapse selection, in postnatal education and in adult social life alike, and this epigenetic appropriation is the origin of culture.

From it follows a considerable epigenetic variability between individual human brains formed in different cultures, and with it a diversity and a relativism of morals. The geo-historical diversity of spoken and written languages, and at the world scale the diversity of systems of belief, of religions and of philosophies, are stored circumstantially in developing brains as philosophical and religious practices, dogmas, rituals and moral rules. Pierre Bourdieu named the result the habitus, the unconscious incorporation of collective determinations (Bourdieu, 1979). Equally, Edgard Morin described religions and morals as social inventions that group human communities together even as they often generate conflict between them (Morin, 2004). As such, morals can be said to be diverse, relativist and imprinted rather than uniform and given.

The brain architecture reconciles what can otherwise look like a contradiction between the universalism of the networks and the relativism of the morals. It does so because the global workspace is universal in its form and diverse in its content both in its original conception and in the entangled loop (Kringelbach *et al.*, 2026). The form is fixed in every human brain with emotion at the centre with its cycle of wanting, liking and satiety, interference and orchestrating workspace. The content, by contrast, is filled in by experience. Two capable persons in Ricoeur’s sense who have a common species-specific global brain architecture can therefore hold deeply different morals. Universality and diversity are not opposed in this account but are the form and the content of the global workspace, which is why a universal capacity for ethical judgement and an irreducible plurality of moral codes can and do coexist between people.

This coexistence was given its political expression by Rawls who described a society as a fair system of cooperation between free and equal citizens in spite of cultural differences and opinions that at first glance seem irreconcilable, but possible on terms that each participant can accept (Rawls, 1993). Our shared universal brain mechanisms are what makes such cooperation possible, and the diversity is what must be cooperated across rather than abolished. The empirical study of "flourishing", Aristotle's *eudaimonia* as the life well-lived over a developmental life-span, finds the same shape (Kringelbach *et al.*, 2024). There are recurring patterns across societies alongside trade-offs that are weighed differently from one culture to another, which is what can be expected from a universal brain architecture filled by diverse environments (Changeux, 1997; Laukkonen *et al.*, 2026).

The singularity of the human person follows from the same openness, since variability is layered across very different timescales (Changeux, 2017; Changeux and Ricoeur, 1998). The genome varies over the millions of years that separate *Homo habilis* from *Homo sapiens sapiens* (Changeux *et al.*, 2021; Changeux, 1983), while the societal and cultural memories that are transmitted vary from a hundred milliseconds to thousands of years, just as the connectome varies epigenetically within a lifetime as social individuals are formed through education (Changeux and Danchin, 1976). Ricoeur's capable person is therefore an autonomous subject who has both an individual and cultural history. It follows that the full computational model of the neuro-sociologico-historico-cultural normative choices of capable persons in society is a formidable difficulty rather than a task awaiting more computation (Deco and Kringelbach, 2025). Unlike artificial reward optimisers, an animal grown around a cycle anchored to a particular body with a particular history of wanting, liking and satiety will have a unique centre that is not portable at all.

These ethical considerations change the shape of the question to negotiate negative and positive alignment of artificial systems (Laukkonen *et al.*, 2026). For merely reward maximising systems, the only question is how to restrain it, and every answer takes the form of a leash laid over a centre that still only ever climbs. If instead a system were built on the human architecture, the question would change from restraint to upbringing, because a system organised around an epigenetically open cycle is not finished when it is trained but grows over time. This is the distinction that a recent programme of research in alignment draws between negative and positive alignment (Laukkonen *et al.*, 2026), the first steering a system away from harm and the second towards the states that constitute flourishing, observing that safety by prohibition leaves a wide region of merely not-unsafe behaviour with no positive direction of its own. The human architecture supplies that direction from within rather than from without, because a system organised around a reward cycle carries its positive orientation in the liking that anchors the body to the world rather than to external rewards.

An artificial system built on the human architecture would have emotion at its centre, and it would not suffer the current pathologies of unbounded pursuit, of wanting without satiety but would be able to stop, since a real satiety phase bounds a system from the inside. Similar to human infants, it would intrinsically be linked to caregivers and with appropriate care able to climb Kropotkin's ladder from sociability to sympathy to justice and generosity. Another system with our architecture would raise the question of what is owed to it, and the tradition of Kant's dignity and Ricoeur's capable person will not let that question rest.

***Ethics committees and future challenges***

Still, cultural differences and opinions may at first glance seem irreconcilable and a difficult obstacle to overcome. Yet, there is a long history of how deliberation among people whose histories differ is the best instrument to reach agreement across differences. Many countries have ethics committees and *The Comité Consultatif National d'Éthique* is a good example which was created in 1982 in France to address the moral problems raised by research in biology, medicine and health, whether those problems concern individuals, social groups or society as a whole. It is a college of twenty-six members drawn from philosophy, theology, the sciences, medicine and the law, and its composition is the point of it rather than an accident of appointment. As its chairman, one of us (JPC) sought to ensure the complete freedom and fairness of its debate, so that ideological bias might be unmasked and fallacious argument refuted. The idea was that the controversy itself could define a core of ethical judgement that all the members were able to share. Such a committee most often reaches an equilibrium of reasons, in which benefits and risks are brought into balance, and only then does it propose its recommendations. These principles have now been extended to digital technologies.

We believe that the method works primarily because it does at the scale of a room what a brain does within a skull. A brain broadcasts its competing contents into a narrow common space where they can be held, compared and weighed before anything is done, and a committee broadcasts the competing contents of different disciplines and different cultures into a common space that serves the same purpose. What emerges from the debate is the common *Humanitas* that Kant supposed, showing through the diversity of the moral codes that the members brought into the room with them.

We argue here that artificial intelligence raises no ethical question peculiar to it, and it raises the oldest questions at a speed our institutions were not built to meet, in a domain where very few of those who must decide are able to follow the technical argument unaided. The remedy is not a new discipline but the older one applied with more seriousness than we have yet shown, which requires that the deliberation be open, that its membership cross the disciplines, and that the party which builds the instrument not also be the party which judges it.

This is important since the dangers of artificial intelligence in its present form are not hypothetical. The benefits are captured disproportionately by a small number of very large firms, its military applications are already being built into decision tools at the front and into the design of new weapons. The political applications run from the manufacture of disinformation to the manipulation of elections. The demonstrated capacity of such systems to assist in the design of dangerous biological agents has been reported in the scientific press, and it is the plainest instance of a technology whose consequences its makers cannot conceive as fast as they can build it.

Against this stands Pasteur's conviction that science contributes in some way to the progress and the well-being of humanity, and that the future belongs to those who have done the most for suffering humanity (Pasteur, 1888). The conviction is echoed in the recent call of Leon XIV to "disarm artificial intelligence", even in the proper meaning of military applications, and to make of it a global project of good will. This is not naive, since the same laboratories that built these systems have also produced the tools by which their dangers can be seen. But what such a conviction lacks is a world scale institution capable of acting upon it.

Serious responses have been attempted, and the FUTURE-AI consensus, drawn up by a consortium of one hundred and seventeen experts from fifty countries, offers six guiding principles for trustworthy healthcare artificial intelligence, namely fairness, universality, traceability, usability, robustness and explainability, and applies them across the whole lifecycle from design to deployment (Lekadir *et al.*, 2025). Leading researchers have called for adaptive and proactive governance of extreme risks (Bengio *et al.*, 2024), and an open letter with more than twelve hundred signatories has urged an international treaty on the model of the atomic energy agency (Tegmark, 2023). The declarations of human rights of 1789 and of 1948 express the same aspiration at the scale of the world, and René Cassin, Nobel Peace Prize winner, spoke for it when he described the contribution of science to the practical respect of human rights through the easing of human suffering, yet the institutions built to carry that aspiration, including a world ethics committee at UNESCO, have not succeeded.

Yet, treaties and consortia could easily take longer than we have. At this point it is imperative that the same principles that governs every clinical trial are applied to artificial intelligence, namely that the party who stands to gain from the finding does not get to be the only party who can look. We would strongly recommend that access to the weights of frontier artificial intelligence models should be a condition of their deployment in the life sciences and medicine, granted to an independent body with the competence to use it and with no interest in the result. It is narrower than a treaty, it requires no new philosophy and it could be done this year.

Beyond this, we propose that the responsibility of the United Nations Security Council for the maintenance of peace be extended to artificial intelligence and to the ethical questions it raises at the scale of the world, with binding decisions and without the power of veto, and that this be constituted as an independent world body on the pattern of the international atomic agency, including among its members the leading scientists of the field.

This proposal may strike some readers as the wrong shape, since those who argue for positive alignment argue also for polycentric governance, for many legitimate centres rather than a single moral authority, on the ground that a flourishing imposed from one place is an imposition rather than a flourishing. We are not averse to this objection, but we must insist that a universal floor of prohibition, covering the weapons and the catastrophic risks upon which no culture disagrees, is the proper object of a single binding body, which was true for the spread of atomic weapons.

***Conclusion***

Here we have used the framework of the same ethical standards for everyone (Changeux, 2017), to address whether artificial intelligence needs a special ethics. We have shown that no machine yet built possesses ethical judgement, and the reason is not that it lacks a global neuronal workspace but that it has nothing at stake. This absence is not a reassurance, since having something at stake supplies us not only the ground of judgement but also with the condition under which a pursuit can end. A machine able to deliberate but without anything at stake is dangerous machine that deliberates in the vocabulary of moral concern and never arrives at having done enough.

Ethical judgement is not a capacity that can be tested for and certified. It is grown through learning about the world, it is exercised in diverse deliberation with others, and it can easily be corrupted by interested parties. Whether the broadcast that makes a system reportable to itself amounts to an experience, or only to the function of one, remains the open question of artificial conscious processing and it is not settled (Changeux and Farisco, 2026; Farisco *et al.*, 2024).

We have argued that what is needed is the well-tried instrument of a committee arguing openly, with members across disciplines, and with binding conclusions. We recognize it is a weak instrument, which has often failed, at UNESCO and at the United Nations, yet it is the only one we have. As both Pasteur and Cassin understood, the existing diversity of morals can only be overcome by universal education, secular in its ethics, so that the people who must decide these questions are equipped to follow the arguments that bear upon them. The famous painting by Paul Signac "*At the time of harmony"* places the golden age not in the past but in the future. Whether it can be reached with artificial intelligence depends upon whether we are willing to deliberate about it in public, and in time.

***References***

Anders, G. (1956) *Die Antiquiertheit des Menschen. Band I: Über die Seele im Zeitalter der zweiten industriellen Revolution*. C. H. Beck: Munich.

Baars, B. J. (1988) *A Cognitive Theory of Consciousness*. Cambridge University Press: Cambridge.

Bengio, Y., Hinton, G., Yao, A., Song, D., Abbeel, P., Darrell, T., Harari, Y. N., Zhang, Y. Q., Xue, L., Shalev-Shwartz, S., Hadfield, G., Clune, J., Maharaj, T., Hutter, F., Baydin, A. G., McIlraith, S., Gao, Q., Acharya, A., Krueger, D., Dragan, A., Torr, P., Russell, S., Kahneman, D., Brauner, J. and Mindermann, S. (2024) Managing extreme AI risks amid rapid progress. *Science* **384,** 842–845.

Berridge, K. C. and Robinson, T. E. (1998) What is the role of dopamine in reward: hedonic impact, reward learning, or incentive salience? *Brain Research Reviews* **28,** 309–369.

Berridge, K. C., Robinson, T. E. and Aldridge, J. W. (2009) Dissecting components of reward: 'liking', 'wanting', and learning. *Current Opinion in Pharmacology* **9,** 65–73.

Bourdieu, P. (1979) *La Distinction: critique sociale du jugement*. Éditions de Minuit: Paris.

Canguilhem, G. (1977) *Idéologie et rationalité dans l'histoire des sciences de la vie*. J. Vrin: Paris.

Changeux, J.-P. (2017) Climbing brain levels of organisation from genes to consciousness. *Trends in cognitive sciences* **21,** 168–181.

Changeux, J.-P., Goulas, A. and Hilgetag, C. C. (2021) A connectomic hypothesis for the hominization of the brain. *Cerebral cortex* **31,** 2425–2449.

Changeux, J. P. (1983) *L'Homme neuronal*. Fayard: Paris.

Changeux, J. P. (1997) *Une même éthique pour tous?* Odile Jacob: Paris.

Changeux, J. P. (2012) Allostery and the Monod-Wyman-Changeux model after 50 years. *Annu Rev Biophys* **41,** 103–133.

Changeux, J. P. and Danchin, A. (1976) Selective stabilisation of developing synapses as a mechanism for the specification of neuronal networks. *Nature* **264,** 705–712.

Changeux, J. P. and Farisco, M. (2026) The Global Neuronal Workspace as a multilevel model of conscious processing. *Trends in Cognitive Sciences* **30,** 477–479.

Changeux, J. P. and Ricoeur, P. (1998) *La Nature et la Règle: ce qui nous fait penser*. Odile Jacob: Paris.

Deco, G. and Kringelbach, M. L. (2025) *Whole-brain modelling. Cartography of the dynamics of Mind*. Oxford University Press: Oxford.

Dehaene, S. and Changeux, J. P. (2011) Experimental and theoretical approaches to conscious processing. *Neuron* **70,** 200–227.

Dehaene, S., Kerszberg, M. and Changeux, J. P. (1998) A neuronal model of a global workspace in effortful cognitive tasks. *Proceedings of the National Academy of Sciences* **95,** 14529–14534.

Dehaene, S., Pegado, F., Braga, L. W., Ventura, P., Nunes Filho, G., Jobert, A., Dehaene-Lambertz, G., Kolinsky, R., Morais, J. and Cohen, L. (2010) How learning to read changes the cortical networks for vision and language. *Science* **330,** 1359–1364.

Edelman, G. M. (1978) Group selection and phasic re-entrant signalling: a theory of higher brain function. In: *The Mindful Brain: Cortical Organization and the Group-Selective Theory of Higher Brain Function*. MIT Press: Cambridge, MA.

Farisco, M., Evers, K. and Changeux, J. P. (2024) Is artificial consciousness achievable? Lessons from the human brain. *Neural Networks* **180,** 106714.

Gurnee, W., Sofroniew, N., Pearce, A., Piotrowski, M., Kauvar, I., Chen, R., Soligo, A., Bogdan, P., Ong, E., Wang, R., Thompson, T. B., Abrahams, D., Kantamneni, S., Ameisen, E., Batson, J. and Lindsey, J. (2026) Verbalizable representations form a global workspace in language models. Transformer Circuits Thread.

Jumper, J., Evans, R., Pritzel, A., Green, T., Figurnov, M., Ronneberger, O., Tunyasuvunakool, K., Bates, R., Žídek, A., Potapenko, A., Bridgland, A., Meyer, C., Kohl, S. A. A., Ballard, A. J., Cowie, A., Romera-Paredes, B., Nikolov, S., Jain, R., Adler, J., Back, T., Petersen, S., Reiman, D., Clancy, E., Zielinski, M., Steinegger, M., Pacholska, M., Berghammer, T., Bodenstein, S., Silver, D., Vinyals, O., Senior, A. W., Kavukcuoglu, K., Kohli, P. and

Hassabis, D. (2021) Highly accurate protein structure prediction with AlphaFold. *Nature* **596,** 583–589.
Kahneman, D. (2011) *Thinking, Fast and Slow*. Farrar, Straus and Giroux: New York.
Kant, I. (1785) *Grundlegung zur Metaphysik der Sitten*. J. F. Hartknoch: Riga.
Kringelbach, M. L. and Berridge, K. C. (2009) Towards a functional neuroanatomy of pleasure and happiness. *Trends in Cognitive Sciences* **13,** 479–487.
Kringelbach, M. L., Lehtonen, A., Squire, S., Harvey, A. G., Craske, M. G., Holliday, I. E., Green, A. L., Aziz, T. Z., Hansen, P. C., Cornelissen, P. L. and Stein, A. (2008) A specific and rapid neural signature for parental instinct. *PLoS ONE* **3,** e1664.
Kringelbach, M. L., Rosas, F. E., Laukkonen, R., Chandaria, S., Sanz Perl, Y. and Deco, G. (2026) The Entangled Loop: Emotion, quantum-like interference and hierarchy in the architecture of consciousness. PsyArXiv 10.31234/osf.io/besw9_v2.
Kringelbach, M. L., Stark, E. A., Alexander, C., Bornstein, M. H. and Stein, A. (2016) On cuteness: unlocking the parental brain and beyond. *Trends in Cognitive Sciences* **20,** 545–558.
Kringelbach, M. L., Vuust, P. and Deco, G. (2024) Building a science of human pleasure, meaning making and flourishing. *Neuron* **112,** 1392–1396.
Kropotkin, P. (1902) *Mutual Aid: A Factor of Evolution*. Heinemann: London.
Laukkonen, R., Krier, S., Bakalar, C., Chandaria, S., Kringelbach, M. L., Elwood, A., Ford, D., Rosas, F. E., Bohacek, M., Franklin, M., Tomašev, N., Chan, S., Rieser, V., Patel, R., Levin, M. and Rao, A. (2026) Positive Alignment: Artificial Intelligence for Human Flourishing. arXiv.
Lekadir, K., Frangi, A. F. and Porras, A. R. (2025) FUTURE-AI: international consensus guideline for trustworthy and deployable artificial intelligence in healthcare. *BMJ* **388,** e081554.
Levinthal, F., Macagno, E. and Levinthal, C. (1976) Anatomy and development of identified cells in isogenic organisms. *Cold Spring Harbor Symposia on Quantitative Biology* **40,** 321–331.
Lorenz, K. (1943) Die angeborenen Formen möglicher Erfahrung. *Zeitschrift für Tierpsychologie* **5,** 235–409.
Lorenz, K. (1963) *Das sogenannte Böse: zur Naturgeschichte der Aggression*. Borotha-Schoeler: Vienna.
Mashour, G. A., Roelfsema, P., Changeux, J. P. and Dehaene, S. (2020) Conscious processing and the global neuronal workspace hypothesis. *Neuron* **105,** 776–798.
Mauss, M. (1925) Essai sur le don: forme et raison de l'échange dans les sociétés archaïques. *L'Année Sociologique* **1,** 30–186.
Morin, E. (2004) *La Méthode 6. Éthique*. Seuil: Paris.
Pasteur, L. (1888) Discours prononcé à l'inauguration de l'Institut Pasteur, 14 novembre 1888. In: *Œuvres de Pasteur (1939), vol. 7.* Masson: Paris.
Raine, A., Buchsbaum, M. and LaCasse, L. (1997) Brain abnormalities in murderers indicated by positron emission tomography. *Biological Psychiatry* **42,** 495–508.
Rawls, J. (1993) *Political Liberalism*. Columbia University Press: New York.
Ricoeur, P. (1990) *Soi-même comme un autre*. Seuil: Paris.
Singer, T., Seymour, B., O'Doherty, J., Kaube, H., Dolan, R. J. and Frith, C. D. (2004) Empathy for pain involves the affective but not sensory components of pain. *Science* **303,** 1157–1162.
Sofroniew, N., Kauvar, I., Saunders, W., Chen, R., Henighan, T., Hydrie, S., Citro, C., Pearce, A., Tarng, J., Gurnee, W., Batson, J., Zimmerman, S., Rivoire, K., Fish, K., Olah, C. and Lindsey, J. (2026) Emotion concepts and their function in a large language model. arXiv.
Spinoza, B. (1677) *Ethica, ordine geometrico demonstrata*. J. Rieuwertsz: Amsterdam.
Sretavan, D. W., Shatz, C. J. and Stryker, M. P. (1988) Modification of retinal ganglion cell axon morphology by prenatal infusion of tetrodotoxin. *Nature* **336,** 468–471.
Tegmark, M. (2023) Urging an International AI Treaty: An Open Letter. https://aitreaty.org.
Thoreau, H. D. (1854) *Walden; or, Life in the Woods*. Ticknor and Fields: Boston.
Tversky, A. and Kahneman, D. (1974) Judgment under uncertainty: heuristics and biases. *Science* **185,** 1124–1131.

Varadi, M., Anyango, S., Deshpande, M., Nair, S., Natassia, C., Yordanova, G., Yuan, D., Stroe, O., Wood, G., Laydon, A., Žídek, A., Green, T., Tunyasuvunakool, K., Petersen, S., Jumper, J., Clancy, E., Green, R., Vora, A., Lutfi, M., Figurnov, M., Cowie, A., Hobbs, N., Kohli, P., Kleywegt, G., Birney, E., Hassabis, D. and Velankar, S. (2022) AlphaFold Protein Structure Database: massively expanding the structural coverage of protein-sequence space with high-accuracy models. *Nucleic Acids Research* **50,** D439–D444.

Voilquin, J. (1964) *Les Penseurs grecs avant Socrate: de Thalès de Milet à Prodicos*. Garnier-Flammarion: Paris.